\documentclass[12pt]{article}

\usepackage{booktabs} 
\usepackage{authblk}
\usepackage[utf8]{inputenc}
\usepackage{hyperref}
\usepackage{soul} 
\usepackage{multirow}
\usepackage{natbib} 
\usepackage{amsthm,amsfonts}
\usepackage{amsmath,amssymb} 
\usepackage{color}
\usepackage{bm}
\usepackage{caption, subcaption, graphicx, adjustbox, rotating}
\usepackage{float}
\newcommand{\blind}{0}
\definecolor{cardinal}{rgb}{0.77, 0.12, 0.23}

\begin{document}
\def\spacingset#1{\renewcommand{\baselinestretch}%
{#1}\small\normalsize} \spacingset{1}

\if0\blind
{
  \title{\bf A review of Design of Experiments courses offered to undergraduate students at American universities}
 \author[1]{Alan R. Vazquez}
 \author[2]{Xiaocong Xuan}
\affil{School of Engineering and Sciences, Tecnologico de Monterrey, Mexico.}
\affil[2]{Department of Statistics, Columbia University, U.S.A.}

  \maketitle
} \fi

\if1\blind
{
  \bigskip
  \bigskip
  \bigskip
  \begin{center}
    {\LARGE\bf A review of Design of Experiments courses offered to undergraduate students at American universities}
\end{center}
  \medskip
} \fi

\bigskip
\begin{abstract}
\noindent Design of Experiments (DoE) is a relevant class to undergraduate students in the sciences, because it teaches them how to plan, conduct, and analyze experiments. In the literature on DoE, there are several contributions to its pedagogy, such as easy-to-use class experiments, virtual experiments, and software to construct experimental designs. However, there are virtually no systematic evaluations of the actual DoE pedagogy. To address this issue, we build the first database of DoE courses offered to undergraduate students in the United States. The database has records on courses offered from 2019 to 2022 at the best universities in the US News Best National Universities ranking of 2022. Specifically, it has data on 18 general and content-specific features of 206 courses. To study the DoE pedagogy, we analyze the database using descriptive statistics and text mining. Based on our analysis, we provide instructors with recommendations and teaching material to enhance their DoE courses. The database and material are included in the supplement of this article. 
\end{abstract} 

\noindent%

\spacingset{1.45} 

\pagebreak

\section{Introduction} \label{sec:introduction}

Experimentation is a core component of the scientific method. Therefore, it is an important learning goal for students in undergraduate programs in the sciences. For example, the American Association of Physics Teachers states that the design of experiments is a skill that must be nurtured in undergraduate physics programs \citep{Kozminski2014}. In its \textit{Levers for Change} report, the American Association for the Advancement of Science identifies that significant experience with experimentation benefits the learning of undergraduate students in chemistry and biochemistry \citep{Laursen2019}. \cite{Cezeaux2006} and \cite{Li2014} show that activities that enhance this experience must be part of undergraduate curricula in biomedical engineering and medicine, respectively. \cite{Estes2002} state that undergraduate programs in civil engineering must demonstrate that graduates have the ability to design and analyze experiments. For all these reasons, it is crucial to teach experimentation well. 

Within the United States (US), undergraduate programs in the sciences teach experimentation to their students through courses about (or that borrow concepts from) the subfield of statistics called Design of Experiments (DoE). DoE concerns statistical tools used to construct experimental designs and analyze their data. Formally, an experimental design is a structured plan to conduct a series of tests of a system (e.g., a process or a product) in which deliberate changes are made to the settings of the system's controllable factors. Using a design, an experimenter collects data on one or more responses, which are analyzed using statistical methods to establish cause-and-effect relationships between them and the controllable factors.  

In the literature on DoE, there are several contributions to its pedagogy. For example, \cite{hunter_ideas_1977} discusses the experience of his students when designing real-life experiments and shows that they improve students' understanding of DoE. \cite{ross_1988} develops a DoE training program for business and industry professionals that teaches the fundamental concepts of DoE and how to use them in practice. \cite{bisgaard_teaching_1991} introduces a short-term course for engineers that has the most widely used experimental designs in the industry. Similarly, \cite{zolman_1999} develops a DoE course for biologists. \cite{AntonyCapone1998} discuss a paper helicopter experiment to teach engineers how to plan experiments with several controllable factors. The experiment can be easily used in a classroom to demonstrate DoE concepts. \cite{lye_2011} and \cite{hamada_experiencing_2018} show other easy-to-use experiments to teach DoE. \cite{goos_teaching_2004} show how to use Excel spreadsheets to teach students how to generate designs with flexible numbers of tests.

Other contributions to DoE pedagogy include the development of virtual experiments in which computer software is used to simulate physical experiments. For example, \cite{schwarz2003online} introduces a virtual experiment to study the effect of four aromas on the time it takes a person to complete a paper maze. The simulator for the experiment allows students to test and compare different experimental designs without performing actual experiments. \cite{darius_virtual_2007}, \cite{Steiner2009}, \cite{DeKetelaere2014} \cite{reis_structured_2017}, and \cite{gramacy2020shiny} present other virtual experiments to teach the design and analysis of experiments.

Despite all these contributions, there are virtually no systematic assessments of the features and content covered in DoE courses. Consequently, the actual DoE pedagogy is unknown. To address this issue, \cite{smucker2022} introduce five teaching profiles that provide information on the content, materials, activities, evaluation methods, and future evolution of a DoE course. Four profiles are from the authors' personal reflections, teaching methods, and research and consulting experience. They vary in terms of the focus of the DoE course, including a traditional course, a modern course with connections to data science, a course that uses case studies and active learning, and a course that leverages algorithms to generate flexible experimental designs.

The fifth teaching profile of \cite{smucker2022} is obtained from a survey of 50 instructors of undergraduate and graduate DoE courses offered by statistics departments within the US and Canada. The survey asked instructors about the type and number of students, learning outcomes, topics covered, textbook and software used, prerequisite courses, teaching methods, and assessment used in their DoE courses. The survey shows that instructors promote the learning of the fundamental principles of experimentation (which are randomization, replication, and blocking) more than the theoretical underpinnings of the DoE methodology. It also shows that DoE courses commonly include the topics: \textit{fractional factorial designs}, \textit{block designs}, and \textit{response surface methodology}. To this end, instructors generally use the textbook entitled ``\textit{Introduction to Design and Analysis of Experiments}'' by \cite{montgomery_design_2013} and the statistical software called R \citep{Rsoftware}. 

The survey of \cite{smucker2022} is the first systematic assessment of the DoE pedagogy. However, it has a small sample size, and so its conclusions may not represent the bulk of DoE courses in the US and Canada well. Moreover, the survey focuses on instructors in statistics departments, but DoE is also part of the curricula of programs offered by engineering and mathematics departments, for example. Therefore, the DoE pedagogy in these departments is still unknown.

In this article, we complement the survey of \cite{smucker2022} to provide a complete overview of the DoE pedagogy in the US. We do this by building the first database of DoE courses offered to undergraduate students at universities within this country. The database has records on the top 100 universities according to the US News Best National Universities ranking, which is one of the most trusted sources of higher education data in the US \citep{Morse2008}. It has data on DoE courses offered by a wide range of departments, including departments of statistics, mathematics, psychology, biostatistics, and engineering. The database has 18 variables that record the university and department that offer the course, the instructor's name, the course code and description, the level of education for registration, the prerequisite courses, the textbook and software used, the evaluation methods, and the academic year and season when the course occurred. Eight of these variables were not included in the survey of \cite{smucker2022}.  

Our database has records on a total of 206 DoE courses that occurred in 2019, 2020, 2021, and 2022. To analyze the data, we use descriptive statistics and text mining \citep{klimberg_fundamentals_2016}. In particular, we are the first to use topic analysis \citep{Giuseppe2018} to identify common topics of a course by analyzing its descriptions shown in online university course catalogs. The main findings of our study are:
\begin{itemize}
    \item Most DoE courses in statistics and mathematics departments are available to freshman, sophomore, junior, and senior students. For engineering departments, a large portion of DoE courses are offered at the graduate level to undergraduate students.
    \item Common prerequisites for DoE courses are probability and statistics, statistical theory, data analysis, and applied statistics.
    \item Most DoE courses in statistics, mathematics, and engineering departments use the textbook of \cite{montgomery_design_2013} and the R software.
    \item The main topics of a DoE course are multifactor designs, randomization restrictions, data analysis, and applications. 
    \item Among the 102 universities that we inspected, 35 of them do not offer an undergraduate DoE course in their departments of statistics, mathematics, and engineering! 
\end{itemize}

Using our knowledge of DoE pedagogy, we provide instructors with two recommendations to enhance their teaching. The first is to teach algoritmically-generated designs \citep{Atkinson2007, Goos2011}, which are attractive alternatives to fractional factorial designs for studying multiple factors. We provide the instructor with topics for an undergraduate DoE course and R demos to cover them in class, which are new to the literature on teaching DoE. The second recommendation is to use virtual experiments to assess student learning. Virtual experiments allow students to gain knowledge and experience in the design and analysis of experiments without performing physical experiments that may be expensive and dangerous. We provide an up-to-date list of virtual experiments that have minimal computing requirements and run on current operating systems. We also introduce a new virtual experiment, concerning the classification performance of the machine learning algorithm called random forest \citep[][ch. 8]{james2021} in terms of its tuning parameters. Our virtual experiment can be accessed for free from a Web browser.

The remainder of the article is organized as follows. In Section~\ref{sec:database}, we introduce our methodology for creating the database of DoE courses. In Section~\ref{sec:analysis}, we pose five research questions about the current DoE pedagogy that we answer by analyzing the database. We present our teaching recommendations and material in Section~\ref{sec:recommendations}. In Section~\ref{sec:conclusion}, we end the article with concluding remarks. Supplementary material for this article includes the database, supplementary sections and files with details of our study, and teaching material to implement our recommendations.  

\section{Methodology} \label{sec:database}

Our methodology for collecting data on the DoE courses offered to undergraduate students has three steps. First, we select a sample of universities in the US that may offer these courses. Second, we define the variables for which we obtain the data. Third, we collect the data by inspecting several official and unofficial online university sources. We present each of these steps separately and then the database. 

\subsection{The sample}

We select the sample of universities using the US News Best Colleges (USNBC) ranking of 2022. This is the oldest college ranking currently published and a widely distributed source of higher education data in the US \citep{Morse2008}. It uses a sum of weighted normalized values across 17 indicators of academic quality to determine the overall score and rank of each school \citep{MorseBrooks2021}. The indicators are graduation and retention rates, social mobility, graduation rate performance, undergraduate academic reputation, among others. \cite{MorseBrooks2021} show the complete list of indicators and weights used in the USNBC ranking for 2022. 


Our sample is the universities in the top 100 of the USNBC ranking within the national universities group \citep{USNews2022}. The sample size is 102 because there are universities that share ranks. We chose the sample for three reasons. First, our goal is to study the state of DoE pedagogy in highly reputable universities in the US. Second, the sample has universities that offer a full range of undergraduate, master, and doctoral programs in the sciences. Third, the sample has Texas A\&M University, Ohio State University, Pennsylvania State University, Michigan State University, Purdue University, and the University of Minnesota, which tend to have high undergraduate enrollments \citep{Mitchell2013,Wood2023}. So, the sample has universities where many students could have potentially taken a DoE course. The sample size is 25.5\% of the total number of national universities in the 2022 rankings. 

Other national rankings such as those of the QS World University Rankings 2022 \citep{QSWUR2022} or the Wall Street Journal/Times Higher Education College Rankings 2022 \citep{THER2022} could have been used to select the sample. However, a close inspection of them showed that at least 65\% of our universities are in their top 100, but in a different order. Therefore, our sample represents the best universities perceived by these rankings to a large extent.

\subsection{The variables}

For each course, we collect data on the variables in Table~\ref{tab:variables}. The table shows their labels, description, type, and values. We classify them into two groups: general and content-specific. The first group reports the general features of the courses and the second group describes their content and details.

The general variables are the university and its rank in the 2022 USNBC ranking, the department that offers the course, the code, the academic year's system, the academic season and year when the course occurred, and the education level for registration. In the database, these variables are labeled RANK, UNIVERSITY, DEPARTMENT, CODE, SYSTEM, SEASON, YEAR, and LEVEL. 

The content-specific variables include the textbook and software used, the name of the instructor, the description, and the methods used to assess student learning. They also include whether an undergraduate course on linear regression, probability and statistics, applied statistics, statistical theory and analysis, or mathematics is a prerequisite; we defer our discussion of these variables to later in this section. The content-specific variables are labeled TEXTBOOK, SOFTWARE, INSTRUCTOR, DESCRIPTION, EVALUATION, LIN\_REG, PROB\_STAT, APP\_STAT, STAT\_THEO\_AN, and MATH.

\begin{table}[]
    \caption{The variables in the database.} \label{tab:variables}
    \centering
    \resizebox{\columnwidth}{!}{%
    \begin{tabular}{llll} \toprule
Label	&	Description	&	Type	&	Values	\\ \midrule
RANK	&	Rank in the 2022 USNBC ranking	&	Numeric	&	Integer	\\
UNIVERSITY	&	Name of the university	&	Character	&	Text	\\
DEPARTMENT	&	Name of the department	&	Categorical	&	Statistics, Mathematics, Engineering, Other	\\
CODE	&	Course code	&	Character	&	Alpha numeric	\\
SYSTEM	&	Academic year's system	&	Categorical	&	Semester, Quarter	\\
SEASON	&	Season when the course occurred 	&	Categorical	&	Fall, Winter, Spring	\\
YEAR	&	Academic year when the course occurred 	&	Numeric	&	Integer	\\
LEVEL	&	Education level for registration	&	Categorical	&	General, Advanced, Graduate, Restricted 	\\
LINEAR\_REG	&	Is linear regression required? 	&	Boolean	&	True, False	\\
PROB\_STAT	&	Is probability and statistics required? 	&	Boolean	&	True, False	\\
STAT\_THEO\_AN	&	Is statistical theory and analysis required? 	&	Boolean	&	True, False	\\
APP\_STAT	&	Is applied statistics required? 	&	Boolean	&	True, False	\\
MATH	&	Is mathematics required? 	&	Boolean	&	True, False	\\
TEXTBOOK	&	Textbook used for the course	&	Character	&	Text	\\
SOFTWARE	&	Software used for the course	&	Character	&	Text	\\
INSTRUCTOR	&	Name of the instructor 	&	Character	&	Text	\\
EVALUATION	&	Evaluation methods	&	Character	&	Text	\\
DESCRIPTION	&	Description in the university's course catalog 	&	Character	&	Text	\\
\bottomrule
    \end{tabular}
    } 
\end{table}

Table~\ref{tab:variables} shows that the variable DEPARTMENT is categorical with five categories. They include ``Statistics,'' ``Mathematics,'' and ``Engineering,'' that emphasize our focus on departments in these fields. This is because their undergraduate programs are quantitative in nature, and so they are more likely to offer a DoE course than other departments. Moreover, the \cite{ASA2014} recommends that statistic minors within these departments have a DoE course. The variable DEPARTMENT also has the category ``Other'' to label departments from other fields.

Another categorical variable is LEVEL with four categories that are described as follows. The category ``General'' refers to a course that can be taken by freshmen, sophomore, junior, and senior students. The category ``Advanced'' refers to a course that can only be taken by junior and senior students. Graduate courses open to undergraduate students fall into the category ``Graduate.'' Lastly, the category ``Restricted'' refers to a course that can only be taken by students in specific major programs, such as statistics or mathematics. 

The variables LINEAR\_REG, PROB\_STAT, APP\_STAT, STAT\_THEO\_AN, and MATH are Boolean and take the value ``True'' or ``False''. They record and categorize the prerequisite courses of a DoE course. The variable LINEAR\_REG takes the value ``True'' if the prerequisite of the DoE course focuses on linear regression, regression analysis, or linear models. Otherwise, it takes the value ``False''. The other Boolean variables work similarly, but with different prerequisites. Specifically, the variable PROB\_STAT equals ``True'' if the DoE course requires a fundamental or introductory course on probability or statistics. The variable STAT\_THEO\_AN is ``True'' if there is a prerequisite on statistical theory and analysis, which includes courses on statistical inference, mathematical statistics, statistical models and methods, and data analysis. The variable APP\_STAT is ``True'' if there is a prerequisite on topics related to the application of statistics, such as applied statistics, statistical practice, biostatistics, engineering statistics, and Bayesian statistics. The variable MATH is ``True'' if the prerequisites are calculus or linear algebra. Two or more of these variables can be ``True'' simultaneously, since a DoE course may have several prerequisites.

The courses surveyed by \cite{smucker2022} provide data on the variables DEPARTMENT, TEXTBOOK, SOFTWARE, and EVALUATION, as well as on prerequisites that we summarize using the variables PROB\_STAT, APP\_STAT, STAT\_THEO\_AN, and MATH. To some extent, they also show data on the variable LEVEL because they were classified into courses taken by undergraduate, master's, or PhD students, or a combination thereof. However, they do not provide data on the variables RANK, UNIVERSITY, CODE, SYSTEM, SEASON, YEAR, INSTRUCTOR, and DESCRIPTION.

\subsection{Data collection process} \label{sec:data_collection}

Our data collection process consists of searching through five online sources: the universities' undergraduate and graduate course catalogs, the educational websites called Course Hero (\url{https://www.coursehero.com/}) and Coursicle (\url{https://www.coursicle.com/}), and the personal website of the instructor. We describe this process below. 

First, we inspect the university's online undergraduate course catalog. Generally, a catalog is either unorganized or organized by department. For an unorganized catalog, we search for the keywords ``doe,'' ``design of experiments,'' ``experiment,'' and ``exp'' in it. If a DoE course is found, we collect data on the variables in Table~\ref{tab:variables} using the course catalog and the website of the department that offers it. 

For an undergraduate course catalog that is organized by department, we limit the search for a DoE course to the departments in the fields of statistics, mathematics, and engineering; with a preference for the former because DoE is a subfield of statistics. We search for the keywords above in the undergraduate course catalog of the statistics department. If a DoE course is found, we collect data on the variables in Table~\ref{tab:variables} using this catalog and the website of this department. In this case, we do not inspect the other departments. Otherwise, we search for the keywords in the undergraduate course catalogs of the mathematics and engineering departments, if available. For all courses found in these departments, we collect data on the variables in Table~\ref{tab:variables}. 
 
The data collection for some variables in Table~\ref{tab:variables} needs additional details. For the variable DEPARTMENT, we assign the department to one of the four categories in the table as follows. Let $C_1$ and $C_2$ denote the category ``Mathematics'' and ``Engineering,'' respectively. We classify the department as $C_i$ if its name contains that of $C_i$. We classify the department as ``Statistics'' if its name has this word, but not any of the names of $C_1$ and $C_2$. If the department does not have the word ``Statistics'' nor any of these category names, we label it as ``Other''. For the variables LINEAR\_REG, PROB\_STAT, STAT\_THEO\_AN, APP\_STAT, and MATH, we first make a list of all prerequisites of the DoE course, and then answer each of the corresponding questions in the table using this list. If a university does not have an undergraduate DoE course, we investigate whether a graduate DoE course is open to undergraduate students according to the official policy and requirements of the university. We classify this type of course as ``Graduate'' in the variable LEVEL. In this case, we collect data for the variables in Table~\ref{tab:variables} using the university's online graduate course catalog. 

For courses in which the course catalogs and the department's website do not provide data on all variables, we inspect Course Hero and Coursicle which provide information on academic courses offered around the world. These websites may show the course description, the instructor's name, the syllabus, the years when the course occurred, among other features. Using the data available from these features, we complete the missing entries of as many variables as possible for those courses. 

If those websites do not provide data on all variables for a course, we inspect another online source. More specifically, for courses in which the name of the instructor is available, we search for the instructor's personal website and, if available, for the syllabus of his or her DoE course. If the syllabus is accessible, we store it and complete as many of the remaining missing variables' entries as possible. 

\subsection{The database}

Our method generated a database with 206 rows and 18 columns. Each row corresponds to a DoE course, and each column corresponds to a variable in Table~\ref{tab:variables}. The database has records of 70 DoE courses with unique course codes. In what follows, we refer to these courses as \textit{unique} DoE courses. The database shows that these courses were offered by 23 departments in 67 universities during the academic years of 2019, 2020, 2021, and 2022. Seven and four departments are in the categories ``Engineering'' and ``Other,'' respectively, and the rest are equally divided into the categories ``Statistics'' and ``Mathematics.'' We have that 42, 12, and 10 of the 70 unique DoE courses are offered by statistics, mathematics, and engineering departments, respectively. The rest are offered by other departments.

The database has courses offered by 67 of the 102 universities in the sample. For the other 35 universities, we did not find a DoE course offered to undergraduate students using our data collection process. Table A1 in Supplementary Section~A lists these universities. They include Columbia University, the Massachusetts Institute of Technology, Stanford University, the University of Pennsylvania, and Georgia Institute of Technology. We conclude that these universities do not have an undergraduate DoE course in their departments of statistics, mathematics, and engineering.

For 26 courses, the database has data on all variables in Table~\ref{tab:variables}. For the other courses, the database misses the values of one or more variables. This was the case for the variables SEASON, YEAR, INSTRUCTOR, TEXTBOOK, EVALUATION, and SOFTWARE, which have a missing value rate in the database of 2.91\%, 3.88\%, 12.62\%, 57.77\%, 84.47\%, and 86.89\%, respectively. All other variables have values for all courses in the database. 

We include an Excel file with the database in the supplementary material of this article. To complement the database, the supplementary material also has 35 syllabi that we obtained using our data collection process.

\section{Data analysis} \label{sec:analysis}

Our database allows us to answer a wide variety of research questions about the pedagogy of DoE. Here, we concentrate on the following questions:
\begin{enumerate}
    \item[Q1:] What is the level of education required to take a DoE course? 
    \item[Q2:] What are the prerequisite courses for a DoE course?
    \item[Q3:] What are the main topics covered in a DoE course?
    \item[Q4:] What are the common textbook(s) and software(s) used in a DoE course?
    \item[Q5:] What are the main evaluation methods used in a DoE course?
\end{enumerate}

In Sections~\ref{sec:Q2} to \ref{sec:Q5}, we answer these questions by analyzing the database, and, in Section~\ref{sec:discussion}, we give a discussion. The analysis in Q1, Q2, and Q3 ignores the iterations of the unique DoE courses that occurred between 2019 and 2022. This is because the data of the variables used in the analysis do not change between years. The analysis in Q4 and Q5 involves all 206 courses in the database, since the data of the variables used vary between years. For Q1, Q2, and Q4, the amount and structure of the data when partitioned by department allow us to comment on DoE courses offered by the departments of statistics, mathematics, and engineering.

\subsection{Answer to research question Q1} \label{sec:Q2}

For the 70 unique DoE courses in the database, we calculated the frequencies of the categories of the variable LEVEL. We have that 38 of these courses can be taken by freshman, sophomore, junior, and senior students, because they fall into the category ``General.'' Moreover, 17 unique courses are in the category ``Graduate'' and so, they are part of a graduate program and open to undergraduate students. Nine and six of the unique courses are in the category ``Advanced'' or ``Restricted.'' In other words, they are available to junior and senior students only or restricted to students in specific major programs.

\begin{figure}[h!]
    \centering
    \includegraphics[width=0.8\textwidth]{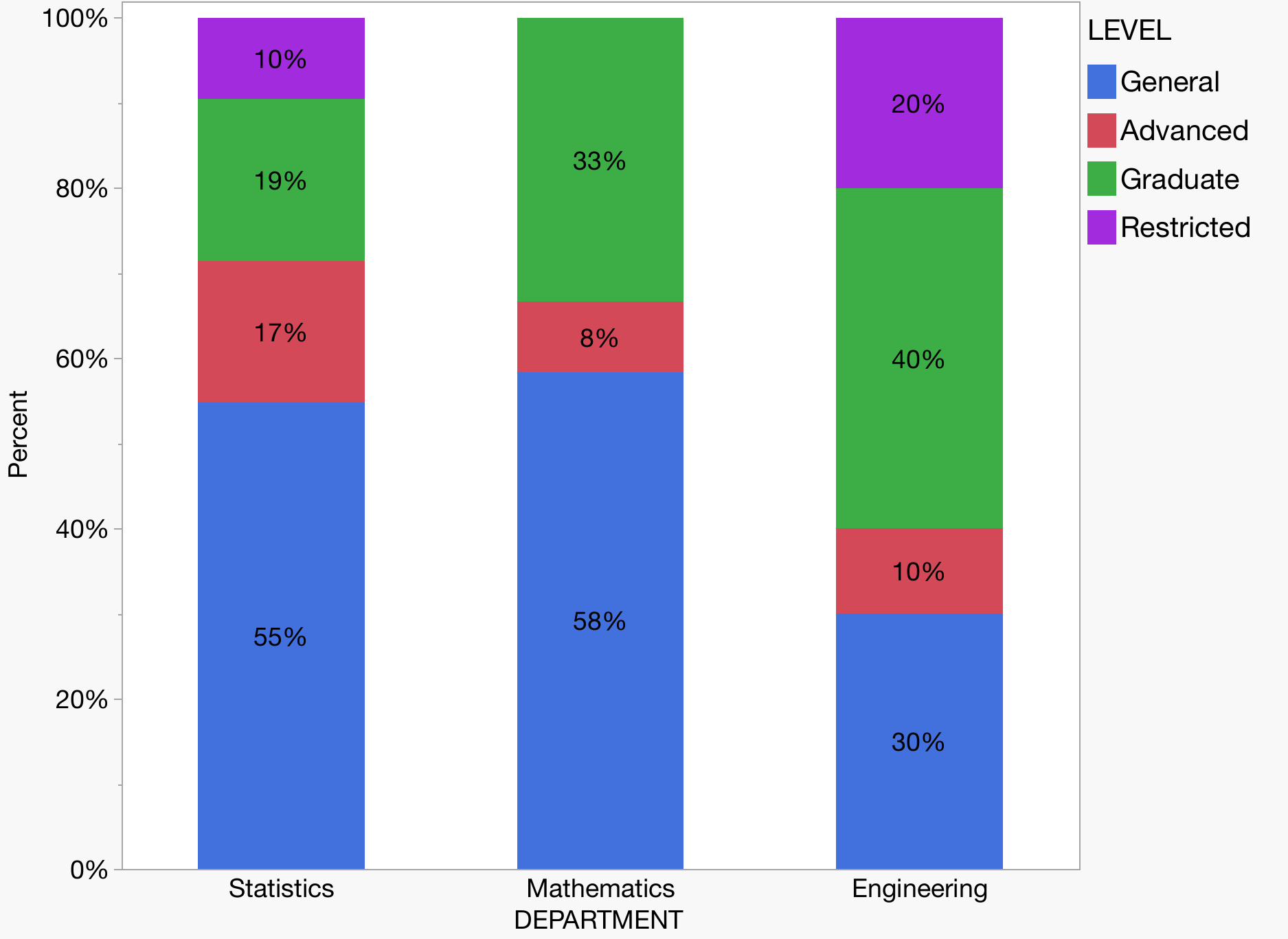}
    \caption{Relative frequency, stacked bar chart for the required education level for the 64 unique DoE courses in the statistics, mathematics, and engineering departments. The online version of this figure is in color.} 
    \label{fig:level_by_department}
\end{figure}

Figure~\ref{fig:level_by_department} shows the level of education required for the 64 unique DoE courses offered by the departments of statistics, mathematics, and engineering. Specifically, it shows a relative frequency, stacked bar chart of the variables LEVEL and DEPARTMENT for them. More than half of these unique DoE courses in the departments of statistics and mathematics are offered to all undergraduate students without restriction or the need for permission, because they are in the category ``General.'' This shows the relevance of a DoE course for undergraduate programs in these departments. In contrast, most unique DoE courses in engineering departments belong to the category ``Graduate'', meaning that undergraduate students generally take DoE at a graduate level there.

\subsection{Answer to research question Q2}

Table~\ref{tab:count_pre-requisites} summarizes the prerequisites for the 70 unique DoE courses in the database. It shows the number of unique DoE courses for which one or two of the variables LINEAR\_REG, PROB\_STAT, STAT\_THEO\_AN, APP\_STAT and MATH are ``True'' simultaneously. Table A2 in Supplementary Section~A lists the names of the prerequisites that fall within the categories given by these variables. 

Table~\ref{tab:count_pre-requisites} shows that a course in statistical theory and analysis is the most common prerequisite because it is required in 24 of the 70 unique courses. The second most common prerequisite is probability and statistics, as 21 of these courses need it. Table~\ref{tab:count_pre-requisites} shows that 15, 20 and 6 of the other unique DoE courses have prerequisites in linear regression, applied statistics, and mathematics, respectively.

Regarding combinations of two prerequisites, Table~\ref{tab:count_pre-requisites} shows that eight of the 70 unique DoE courses require students to take courses in both probability and statistics, and statistical theory and analysis. The same is true for the pair of prerequisites involving applied statistics and statistical theory and analysis. Five or fewer of the unique DoE courses need other combinations of two prerequisites. Although not shown in Table~\ref{tab:count_pre-requisites}, combinations of three or more prerequisites are rarely needed.

\begin{table}[h!]
    \caption{Count of unique DoE courses for which one or two of the variables LINEAR\_REG, PROB\_STAT, STAT\_THEO\_AN, APP\_STAT and MATH are ``True'' simultaneously.}
    \centering
    \begin{tabular}{lc} \toprule
    Variable(s) & Count \\ \midrule 
LINEAR\_REG		&					15 \\ 
PROB\_STAT		&					21 \\
STAT\_THEO\_AN	&						24 \\
APP\_STAT	&						20 \\
MATH			&				6 \\
LINEAR\_REG	and	PROB\_STAT		&			2 \\
LINEAR\_REG	and	STAT\_THEO\_AN		&			3 \\
LINEAR\_REG	and	APP\_STAT		&			2 \\
LINEAR\_REG	and	MATH		&			0 \\
PROB\_STAT	and	STAT\_THEO\_AN		&			8 \\
PROB\_STAT	and	APP\_STAT		&			5 \\
PROB\_STAT	and	MATH			&		1 \\
STAT\_THEO\_AN	and	APP\_STAT	&				8 \\
STAT\_THEO\_AN	and	MATH	&				4 \\
APP\_STAT	and	MATH	&				1 \\
\bottomrule
    \end{tabular} 
    \label{tab:count_pre-requisites}
\end{table} 

To study prerequisites by department, we concentrate on courses in probability and statistics, statistical theory and analysis, and applied statistics because they have the highest frequencies in Table~\ref{tab:count_pre-requisites}. Table~\ref{tab:prerequisities_by_department} shows the percentage of unique DoE courses that require each type of prerequisite for the departments of statistics, mathematics, and engineering. For example, the last row of the table involves the numbers of courses offered by engineering departments for which PROB\_STAT, STAT\_THEO\_AN, and APP\_STAT is ``True''. We calculate the percentage in the table by dividing these numbers by 10, the total number of unique DoE courses offered in these departments. The percentages in each row do not add up to 100\% because a course may have two or more of these prerequisites. 

Table~\ref{tab:prerequisities_by_department} shows that 42.8\% of the unique DoE courses offered in statistics departments need a course in statistical theory and analysis. This percentage is higher than those of the other prerequisites. We have that 33.3\% of the unique DoE courses offered in mathematics departments require a course in that topic or applied statistics. This percentage is higher than that of probability and statistics for these departments. We have that 30\% of the unique DoE courses offered in engineering departments require applied statistics and probability and statistics, and 20\% need a course in statistical theory and analysis. 

\begin{table}[h!]
    \caption{Percentages of unique DoE courses for which PROB\_STAT, STAT\_THEO\_AN, and APP\_STAT is ``True'' by department.}
    \centering
    \begin{tabular}{cccc} \toprule
               & \multicolumn{3}{c}{Prerequisite} \\ \cmidrule{2-4}
    DEPARTMENT & PROB\_STAT & STAT\_THEO\_AN & APP\_STAT\\ \midrule 
Statistics		&					35.7 & 42.8  & 30.9 \\ 
Mathematics		&					25.0  &   33.3 &  33.3 \\
Engineering	&						30.0  &   20.0 &  30.0 \\
\bottomrule
    \end{tabular} 
    \label{tab:prerequisities_by_department}
\end{table} 

We also analyzed combinations of prerequisites by department. Around 14\% of the courses offered in statistics departments need statistical theory and analysis in combination with either probability and statistics, or applied statistics. This percentage is twice as high than for the pair involving applied statistics and probability and statistics. For mathematics and engineering departments, combinations of two prerequisites are rarely needed. 

\subsection{Answer to research question Q3}
\label{sec:Q6}

To identify the main topics of a DoE course, we analyze the descriptions of the 70 unique DoE courses using the text mining technique called topic analysis \citep{Giuseppe2018}. Topic analysis allows us to identify common topics among a large collection of documents. It is similar to an exploratory factor analysis in multivariate statistics \citep{johnson1992applied} because a topic is a latent factor derived from the usage of particular terms in documents. We refer to \cite{evangelopoulos2013latent} and \cite{Karletal2015} for comprehensive tutorials on topic analysis. 

To conduct our topic analysis, we have to provide technical details about turning the course descriptions, contained within the variable DESCRIPTION in the database, into numerical data. We also need to motivate some of our decisions for the exploratory factor analysis. To save space, here we restrict to describing the two main topics that we found from the analysis. We give details of our topic analysis in the Supplementary Section~B. The supplementary material has a JMP file to reproduce it.

The two main topics of the 70 unique DoE courses are called ``Multifactor Designs With and Without Randomization Restrictions'' and ``Data Analysis and Applications.'' We derived these topics from the first two latent factors obtained from the topic analysis. They account for 27.49\% of the variability in the numerical data generated from the course descriptions; see Supplementary Section~B. The concepts surrounding the first topic are completely randomized, fractional factorial, blocked, split-plot, and nested designs. These concepts are derived from words in the descriptions that appear in the latent factor associated with this topic with a high loading; see Supplementary Section~B.  
 
The concepts surrounding the second topic include statistical methods and models for data analysis, which are derived from words that have a high loading on the latent factor of this topic. The other words that appear in this latent factor suggest that these concepts are explained at an introductory level. They also show that scientific and engineering applications of these concepts are relevant to the unique DoE courses. We refer to Supplementary Section~B for a more technical discussion of these topics and concepts.

\subsection{Answer to research question Q4}
\label{sec:Q4}

We now consider all 206 DoE courses in the database to answer this and the next question. Here, we first discuss the textbooks and then the software used in the DoE courses. Our analysis of textbooks is based on 87 courses that report their use in the database. In other words, it involves the complete entries of the variable TEXTBOOK. Table~\ref{tab:textbooks} shows a frequency table for the textbooks used in these 87 courses. The textbooks were used in the iterations of 33 unique DoE courses taught by different instructors in the academic years of 2019 to 2022. They are published by an educational publishing house, except for \cite{Oehlert2010}, \cite{Yakir2011}, and \cite{BransonSeltman2020} that were written and self-published by DoE instructors. Electronic copies of these textbooks are available for free on the Web. 

Table~\ref{tab:textbooks} shows that the most widely used textbook is an edition of \cite{montgomery_design_2013}, which is used in 40 of the 87 courses. The second most widely used textbook is an edition of \cite{kutner_applied_2005} because it is used in 12 of these courses. The other textbooks are used in six or fewer courses. An analysis of textbooks by department (not shown here) shows that \cite{montgomery_design_2013} is a predominant reference for DoE courses across all. 

The database has 27 courses that report the use of software, given by the complete entries of the variable SOFTWARE. They are iterations of 13 unique courses in different years. By far, the most popular software is R because it is used in 20 of the 27 courses. The second most popular software is JMP, which is used in six of these courses. The rest of these courses use Minitab, SAS, G-Power, Matlab, SPSS, STATA, Excel, or a combination of these. In general, these conclusions hold when analyzing the use of software by department. Therefore, R is the predominant software for teaching DoE.

\begin{table}[h!]
    \caption{Frequency table for the textbooks used in the DoE courses in the database.} \label{tab:textbooks}
    \centering
    \resizebox{\columnwidth}{!}{%
\begin{tabular}{cll} \toprule
Frequency	&	Title	& Reference	\\ \midrule
40	&	\it ``Design and Analysis of Experiments''	&	\cite{montgomery_design_2013}	\\
12	&	\it ``Applied Linear Statistical Models''	&	\cite{kutner_applied_2005}	\\
6	&	\it ``Statistics for Experimenters: Design, Innovation, and Discovery''	&	\cite{box2005statistics}	\\
4	&	\it ``Introduction to Design and Analysis of Experiments''	&	\cite{cobb2008}	\\
3	&	\it ``A First Course in Design and Analysis of Experiments''	&	\cite{Oehlert2010}	\\
3	&	\it ``Experimental Design and Analysis''	&	\cite{BransonSeltman2020}	\\
3	&	\it ``Experiments: Planning, Analysis, and Optimization''	&	\cite{wu2009}	\\
3	&	\it ``Response Surface Methodology''	&	\cite{myers2009}	\\
3	&	\it ``Visualizing Linear Models''	&	\cite{Brinda2021}	\\
2	&	\it ``Designing Experiments for the Social Sciences''	&	\cite{coleman_designing_2019}	\\
2	&	\it ``Fundamental Concepts in the Design of Experiments''	&	\cite{hicks1999}	\\
2	&	\it ``Sampling: Design and Analysis''	&	\cite{Lohr2021}	\\
1	&	\it ``Analysis of Messy Data Volume 1: Designed Experiments''	& \cite{Milliken2009}		\\
1	&	\it ``Applied Statistics and Probability for Engineers''	&	\cite{Montgomert2013}	\\
1	&	\it ``Design and Analysis of Experiments''	&	\cite{dean1999design}	\\ 
1	&	\it ``Introduction to Statistical Thinking (with R, without Calculus)''	&	\cite{Yakir2011}	\\ \bottomrule
\end{tabular}
} 
\end{table} 

\subsection{Answer to research question Q5}
\label{sec:Q5}

The database has records on the evaluation methods for 32 courses given by the complete entries of the variable EVALUATION. These courses are one or more iterations of 14 unique DoE courses in the academic years of 2019 to 2022. Table~\ref{tab:assessment} shows the evaluation tools, their frequency, and a summary of their use in the 32 courses. Homework is a common tool for assessing student learning because almost all courses use them. Generally, they are assigned on a regular basis (e.g., weekly or biweekly) and must be completed individually. 

\begin{table}[h!]
    \caption{Evaluation tools, their frequency, and summary of their use in 32 DoE courses that have records on evaluation methods.}
    \centering
    \begin{tabular}{lll} \toprule
Tool	&	Count & Summary	\\ \midrule 
Homework	&	31 & Individual homework generally assigned on a regular basis.  	\\
\multirow{2}{*}{Exam}	&	\multirow{2}{*}{28} & Intermediate and final exams that are written, individual,	\\
&	 & and in-class.	\\
\multirow{2}{*}{Project}	&	\multirow{2}{*}{20} & One or more projects performed by groups of students	\\
	&	 &  and mostly used as the last assessment.	\\
Quizzes	&	10 & Short weekly individual quizzes.	\\
\multirow{2}{*}{Class participation}	&	\multirow{2}{*}{5} & Attend class, work on problems and quizzes, and 	\\
&	 &  discuss class material.	\\
Paper discussion	&	2 & Reflection papers.	\\
\bottomrule
    \end{tabular} 
    \label{tab:assessment}
\end{table} 

Table~\ref{tab:assessment} shows that other common evaluation tools are exams and projects, since 20 or more of the 32 courses use them. Exams are administered during or at the end of the course. They occur during class and are performed individually in writing. Few courses report their content. For those that do, we found that the exams have multiple-choice or short-answer questions that ask students to evaluate an experimental design or interpret the results of a data analysis conducted using a specific software.

Generally, projects are performed by teams of students and are used as the final assessment of the course. They require students to design and conduct an experiment and analyze its data. Most projects involve physical experiments that are defined by the students or based on a case study obtained from a scientific paper. In a few cases, they involve virtual experiments such as the Island application \citep{bulmer2011life}. Other projects ask students to analyze data obtained on their own, from a textbook, or from the instructor. The project evaluation is done using a written report, an in-class presentation, or both.     

Other evaluation tools for DoE courses are quizzes, class participation, and paper discussions. The quizzes are shorter than an exam and must be completed individually by the students. Class participation includes attending lectures, interacting with the instructor or other students through discussions of class material, and individually working on problems and quizzes. In a paper discussion, students write a reflection paper whose goal is to increase their understanding of the course material. There, they synthesize and criticize the course readings assigned by the instructor. 

\subsection{Discussion} \label{sec:discussion}

Our study of the required level of education and prerequisites for a DoE course allows us to determine the type of student who takes it. Generally, the student is a freshman, sophomore, junior, or senior, and possesses knowledge of probability and statistics, applied statistics, or statistical theory and analysis. Therefore, the student knows the core topics of probability and statistics such as random variables, probability distributions, summary statistics, parameter estimators, the central limit theorem, confidence intervals, and hypothesis tests \citep{Montgomert2013}. 

Interestingly, during the data collection process, we noticed that many of the prerequisites in probability and statistics, applied statistics, and statistical theory and analysis cover linear regression to some extent. This was especially true for courses in statistical theory and analysis and applied statistics, since linear regression is one of the main tools to study the relationship between multiple variables. In fact, 62.85\% of the unique DoE courses have a prerequisite course that features that subject or have it in their topics. Therefore, students typically know about linear models, least-squares estimation, prediction intervals, variance-covariance matrix, and residual analysis \citep{kutner_applied_2005}. Of course, we expect a student who took a full course in linear regression to possess a more in-depth knowledge of these concepts than a student who took other prerequisites.

\section{Recommendations} \label{sec:recommendations}

Based on the results of our data analysis in Section~\ref{sec:analysis}, we make two recommendations to the instructor of a DoE course. They concern teaching alternative experimental designs and using virtual experiments available in the literature.   

\subsection{Recommendation 1: Teach algorithmically-generated designs}

If students have knowledge of linear regression, we recommend the instructor to teach algorithmically-generated experimental designs \citep[also called optimal experimental designs;][]{Atkinson2007, Goos2011}, so that they can tackle more complex experimental situations than it is possible with fractional factorial designs. Algorithmically-generated designs have flexible run sizes, can study multiple types of factors (e.g., categorical, continuous, and blocking) simultaneously, and can accommodate constraints in the experimental design region. This is because they are generated from scratch using computationally efficient algorithms, such as point- and coordinate-exchange algorithms \citep{Cook1980, Meyer1995}. In contrast, fractional factorial designs are typically obtained from catalogs available in texbooks, have run sizes that must be a power of two or three, and cannot include constraints in the experimental region. Therefore, algorithmically-generated designs are more attractive for practical experiments than fractional factorial designs, which were identified as a relevant concept in the DoE courses in Section~\ref{sec:Q6}. Despite their attractive features, however, algorithmically-generated designs are rarely taught in an undergraduate DoE course. Evidence of this is that only three of the 70 unique DoE course descriptions in the database mention them.

Students with knowledge of linear regression have the tools to learn the main concepts and applications of algorithmically-generated designs. For example, their performance for an experiment is evaluated using statistical criteria, which involve the variance-covariance matrix for the least-squares estimates of the coefficients in a user-specified linear regression model. In fact, the standard criterion called D-optimality is the determinant of this matrix \citep{Atkinson2007}. Small values of this criterion imply a small variance in the estimates of the coefficients in the linear regression model. Since students with a background in linear regression know about linear models, least-squares estimates, and the variance-covariance matrix, they will be able to evaluate designs using this criterion. 

To help instructors teach algorithmically-generated designs in an undergraduate DoE course, we provide the following guidance and teaching materials:
\begin{itemize}

    \item Following the teaching profile of Prof. Peter Goos in \cite{smucker2022}, we recommend the instructor teach D-optimal designs for screening and follow-up experiments, I-optimal designs for response surface optimization, D-optimal designs in blocks, and, if time permits, D-optimal split-plot designs. In this way, the instructor can address the topic ``Multifactor Designs With and Without Randomization Restrictions'' identified in Section~\ref{sec:Q6}. Table~\ref{tab:optimaldesign} shows the proposed content along with the names of demos (as supplementary R Markdown files) that we prepared to illustrate the construction, evaluation, and analysis of algorithmically-generated designs. All demos are inspired by real experimental problems that are solved using these designs, except for the first demo in the table that is intended as an introduction. Four demos have real data and illustrate their analysis, so that the instructor can cover the topic ``Data Analysis and Applications'' identified in Section~\ref{sec:Q6}. 

    \item To generate the designs, we recommend the R software and the package called \verb!AlgDesign! \citep{AlgDesign}, which is used in all the demos in Table~\ref{tab:optimaldesign}. We chose the R software because it is the leading software for teaching DoE, as shown in Section~\ref{sec:Q4}. Moreover, it is available for free and, through the \verb!AlgDesign! package, can generate D-optimal designs with or without blocks, I-optimal designs, and D-optimal split-plot designs using the point-exchange algorithm \citep{Cook1980}.
    
    \item The last column of Table~\ref{tab:optimaldesign} shows background readings given by chapters of the textbook entitled ``Optimal design of experiments: A case study approach'' by \cite{Goos2011}. In our experience, this textbook is a great resource for introducing algorithmically-generated designs to undergraduate students. Each chapter begins with a hypothetical case study involving two consultants and a consultee. The consultee faces an experimental problem which the consultants solve by constructing an algorithmically-generated design. Moreover, the consultants show the consultee how to evaluate the design and illustrate its benefits when compared to a standard design. The technical details behind the concepts of each chapter are discussed in a separate section called ``Peek into the black box.'' In this way, the textbook provides a good balance between the application and technicalities of the designs. We refer to \cite{Jensen2012} for a complete review of the textbook. 

    \item We recommend introducing algorithmically-generated designs after two-level fractional factorial designs. This is because their flexibility in terms of their run size is relevant for multifactor experiments and two-level fractional factorial designs are the typical point of entry into these experiments; see the textbook of \cite{montgomery_design_2013}, which we identified as the leading textbook for teaching DoE in Section~\ref{sec:Q4}.
     
\end{itemize}

\begin{table}[h!]
    \caption{Topics and resources to teach algorithmically-generated designs in an undergraduate DoE course. The chapters are from the textbook entitled ``Optimal Design of Experiments: A case study approach'' by \cite{Goos2011}.}
    \centering
    \resizebox{\columnwidth}{!}{%
    \begin{tabular}{lllc} \toprule
Topic	&	Title	&	Demo File	&	Chapter	\\ \midrule
1	&	Introduction to algorithmically-generated designs	&	``Introduction.Rmd''	&	2	\\
2	&	D-optimal designs for screening experiments	&	``Doptimal.Rmd''	&	2	\\
3	&	Algorithmic augmentation of designs	&	``Augment.Rmd''	&	3	\\
4	&	I-optimal designs for response optimization	&	``Ioptimal.Rmd''	&	4	\\
5	&	Constructing a D-optimal design with blocks	&	``Blocked-Designs.Rmd''	&	8	\\
6	&	A D-optimal design with a hard-to-change factor	&	``Split-Plot-Designs.Rmd''	&	10	\\ \bottomrule
    \end{tabular}} 
    \label{tab:optimaldesign}
\end{table}

\subsection{Recommendation 2: Use virtual experiments to evaluate student learning}

Our answer to Q5 shows that projects are common evaluation tools of the DoE courses in the database. Although the design, execution, and analysis of a physical experiment may comprise a good project, it could be time consuming, resource intensive, and potentially dangerous. To avoid these complications, we recommend that instructors use virtual experiments to assess the students' learning of the design and analysis of experiments. 

Table~\ref{tab:simulators} shows six recommended multimedia applications that implement virtual experiments. The table shows the name of the application, the number of factors and responses in its virtual experiment, its computing requirements, and its reference with details of its usage in class. Inspired by the experiment of \cite{antony2002training}, the catapult application simulates the in-flight distance of a ball that is shot using a catapult with five adjustable settings. The Watfactory application of \cite{Steiner2009} simulates a manufacturing process that produces automobile camshafts. It can be used to conduct experiments with the goal of minimizing the variation in the straightness of the camshafts using up to 90 factors. The garden sprinkler application of \cite{DeKetelaere2014} simulates the covered range, the rotation speed, and the water consumption of a garden sprinkler with eight design parameters. The Netflix application introduced in the teaching profile of Prof. Nathaniel Stevens in \cite{smucker2022} simulates the browsing time required by a user to select a movie on the Netflix streaming service. This application simulates different user interfaces given by the settings of four factors. The yield application of \cite{gramacy2020shiny} simulates the yield of a crop plot as a function of the levels of six nutrients. 

\begin{table}[h!]
    \caption{Recommended multimedia applications of virtual experiments.}
    \begin{centering}
    \resizebox{\columnwidth}{!}{%
    \begin{tabular}{lcclll} \toprule
Application	&	Factors	&	Responses	&	Requirements & Reference  \\ \midrule
Catapult$^a$ & 5 & 1  & \multirow{5}{*}{Web Browser} & \cite{antony2002training}  \\
Watfactory$^b$ & 90 & 1 &  & \cite{Steiner2009}   \\
Garden Sprinkler$^c$ & 8 & 3 &  & \cite{DeKetelaere2014}   \\
Netflix$^d$ & 4 & 1  &  & \cite{smucker2022}  \\
Parameter Tuning$^e$ & 7 & 3  &  & This article  \\
Yield & 7 & 1  & R/R Studio & \cite{gramacy2020shiny}  \\
\bottomrule
    \end{tabular}} 
    \end{centering}
\footnotesize{$^a$ \url{https://sigmazone.com/catapult/}} \\ 
\footnotesize{$^b$ \url{https://www.student.math.uwaterloo.ca/~watfacto/login.htm}} \\ 
\footnotesize{$^c$ \url{https://twilights.be/sprinkler/}} \\ 
\footnotesize{$^d$ \url{https://nathaniel-t-stevens.shinyapps.io/Netflix_Simulator_v2/}} \\
\footnotesize{$^e$ 
\if0\blind
{
\url{https://alanrvazquez.shinyapps.io/Parameter_Tuning_RF/}} 
 \else{[BLINDED URL]}
\fi 
}
    \label{tab:simulators}
\end{table} 

The parameter tuning application in Table~\ref{tab:simulators} is a contribution of this paper and inspired by \cite{lujan2018design}, who tuned the (hyper)parameters of random forest \citep[][ch. 8]{james2021} using designed experiments. Specifically, our application simulates the performance of a random forest on a binary classification task in terms of seven of its tuning parameters. The response under study is an estimate of the classification accuracy obtained using 10-fold cross-validation \citep[][ch. 5]{james2021}. There are three classification tasks that involve the detection of three different diseases in a subject. The description of the design problem in the application has been written to minimize the technical details associated with random forest, so that students focus on the design and analysis of experiments to study its tuning parameters. 

The applications in Table~\ref{tab:simulators} have several attractive features for a DoE course. Compared to \cite{schwarz2003online}, \cite{darius_virtual_2007}, and \cite{reis_structured_2017}, they have minimal computing requirements, run on current operating systems, and are available for free. For example, the catapult, Watfactory, garden sprinkler, Netflix, and parameter tuning applications can be accessed through a Web browser without installing software. Although R and R Studio are needed for the yield application, these software have all the required functions to deploy it. Another attractive feature is that the applications allow students to study the effect of up to 90 factors on one or more responses. Therefore, they promote the learning of the topic ``Multifactor Designs With and Without Randomization Restrictions'' identified in Section~\ref{sec:Q6}. Despite these attractive features, none of the DoE courses in the database uses an application in Table~\ref{tab:simulators}. 

The references in Table~\ref{tab:simulators} provide several potential projects involving the virtual experiments. Here, we present a project for the parameter tuning application that the first author has used successfully in his undergraduate DoE course. The goal of the project is to identify the influential tuning parameters on the 10-fold cross-validated estimate of accuracy for a classification task selected by the instructor beforehand. To achieve this goal, students are given a maximum number of runs that they can perform. They must compare two-level fractional factorial designs and D-optimal designs with different run sizes to select a good design for the experiment. Next, they must perform the selected design using the application, collect the response values for the experimental tests, and analyze the data to identify the influential factors. A challenge for them is to accommodate an appropriate number of runs for the main experiment, while accounting for runs to conduct follow-up experiments to resolve ambiguities in the data analysis (if any) or confirmatory experiments to validate their conclusions. Project guidelines are given to students with questions that they must answer to achieve the goal. The supplementary material contains a template with these questions. 

In the first author's experience, the project can be completed in one week by groups of two or three students. To grade the project, the first author has used a written report, following most DoE courses in the database; see Section~\ref{sec:Q5}. We obtain alternative versions of the project using a different maximum number of runs and classification task. 

\section{Conclusion} \label{sec:conclusion}

In this article, we introduced the first database of courses in Design of Experiments offered to undergraduate students in the US. Our database has records of courses offered by several departments within the top 100 universities in the 2022 US News Best National Colleges ranking. Our main data sources were the university's online undergraduate and graduate course catalogs, the educational websites called Coursehero and Coursicle, and the personal websites of the instructors. Using the database, we studied several aspects of DoE pedagogy, such as the required education level, prerequisite courses, textbook and software used, assessment tools, and course topics. To this end, we used descriptive statistics and text mining. Based on our analysis of the database, we provided instructors with two recommendations and teaching material to enhance their DoE courses. 

Our study of DoE pedagogy complements that of \cite{smucker2022} by providing a statistical analysis of data from DoE courses that we obtained from official university sources, instead of a survey of DoE instructors. Papers like these are hopefully the ﬁrst in an ongoing evaluation of DoE pedagogy that tracks its evolution in the future. In recent years, new DoE theory and methods have been developed to solve problems in data science. For example, DoE-inspired subsampling methods have been introduced to reduce big data sets \citep{Wangetal2019,Wangetal2021,ShiTang2021}. Moreover, online controlled experiments are currently being used in high-tech companies such as Google, Netflix, and Facebook \citep{gupta2019top,luca2021power}, and new design and analysis methods are being developed for them \citep{yang2017framework,StevensHagar2021,Shietal2022,Larsenetal2023}. It will be interesting to see if these developments at the interface of DoE and data science gain ground in undergraduate DoE courses. Only another assessment of the DoE pedagogy in the next few years will tell.


\begin{center}
{\large\bf Supplementary Material}
\end{center}

\begin{itemize}

\item \textbf{Supplementary sections.pdf}. Document with additional details on the study of the DoE pedagogy.
\item \textbf{Database.xlsx}. Excel file with (1) the database of 206 DoE courses, (2) the description of the 70 unique DoE courses, and (3) the classification of departments into statistics, mathematics, engineering, and other.
\item \textbf{Syllabi.zip}. Zip file containing the pdf files of 35 syllabi of DoE courses.
\item \textbf{Topic Analysis.jmp}. JMP file to reproduce the topic analysis of course descriptions.
\item \textbf{Material for Algorithmically-Generated Designs.zip}. Zip file containing R markdown files with the demos to teach algorithmically-generated designs and a word document with the sample task description for the parameter tuning application.
\end{itemize}

\if0\blind
{
\bigskip
\begin{center}
{\large\bf Acknowledgments}
\end{center}

The research that led to this article started as part of the undergraduate course ``STATS 199: Directed Research in Satistics'' at the University of California, Los Angeles. The first author thanks his teaching mentors, Peter Goos, Robert Gould, and Maria Cha, who have helped him craft his DoE teaching style.
}\fi

\bibliographystyle{apalike} 
\bibliography{bibliography} 

\end{document}